\documentclass[12pt]{article}
\pdfoutput=1

\usepackage{scicite}

\usepackage{times}

\usepackage{amsmath}
\usepackage{amsfonts}
\usepackage{amssymb}
\usepackage{graphicx}

\usepackage{float}
\usepackage{physics}
\usepackage{multirow}
\usepackage{ulem}
\usepackage{soul}
\usepackage{siunitx}

\DeclareSIUnit{\Molar}{M}

\newenvironment{sciabstract}{%
\begin{quote} \bf}
{\end{quote}}

\usepackage[dvipsnames]{xcolor}
\usepackage{array}
\usepackage{booktabs}
\usepackage{afterpage}
\usepackage{pdfpages}

\makeatletter
\AtBeginDocument{\let\LS@rot\@undefined}
\makeatother

\title{Large-Enhancement Nanoscale Dynamic Nuclear Polarization Near a Silicon Nanowire Surface} 

\author
{Sahand Tabatabaei,${}^{1,2\dagger}$ Pritam Priyadarsi,${}^{1,2\dagger}$ Namanish Singh,${}^{1,2}$\\ Pardis Sahafi,${}^{1,2}$ Daniel Tay,${}^{1,2}$ Andrew Jordan,${}^{1,2}$ Raffi Budakian${}^{1,2\ast}$\\
\\
\normalsize{${}^{1}$Department of Physics and Astronomy, University of Waterloo, Waterloo, ON, Canada, N2L3G1}\\
\normalsize{${}^{2}$Institute for Quantum Computing, University of Waterloo, Waterloo, ON, Canada, N2L3G1}\\
\\
\normalsize{$^\ast$To whom correspondence should be addressed; E-mail:   rbudakian@uwaterloo.ca.}\\
\small{$^\dagger$ These authors contributed equally to this work.}
}

\date{}

\begin{document} 

 \baselineskip24pt

 \maketitle

\begin{sciabstract}
Dynamic nuclear polarization (DNP) has revolutionized the field of NMR spectroscopy, expanding its reach and capabilities to investigate diverse materials, biomolecules, and complex dynamic processes. Bringing high-efficiency DNP to the nanometer scale would open new avenues for studying nanoscale nuclear spin ensembles, such as single biomolecules, virus particles, and condensed matter systems.
Combining pulsed DNP with nanoscale force-detected magnetic resonance measurements, we demonstrated a 100-fold enhancement in the Boltzmann polarization of proton spins in nanoscale sugar droplets at 6~K and 0.33~T.
Crucially, this enhancement corresponds to a factor of 200 reduction in the averaging time compared to measurements that rely on the detection of statistical fluctuations in nanoscale nuclear spin ensembles. These results significantly advance the capabilities of force-detected magnetic resonance detection as a practical tool for nanoscale imaging.

\end{sciabstract}

\section*{Introduction}

Nuclear magnetic resonance (NMR) plays a pivotal role in modern science and technology, with applications ranging from characterizing molecular structure and dynamics, to studying complex systems in condensed matter physics and advancing diagnostic imaging in medicine. In recent decades, extensive research has gone into bringing the powerful capabilities of magnetic resonance, particularly magnetic resonance imaging (MRI), to the nanometer scale \cite{Degen2009,Nichol2012,Maletinsky2012,Nichol2013,Haberle2015,Kost2015,Ajoy2015,Rugar2015,Wang2019,Rose2018,Abobeih2019,Grob2019,Haas2021,Krass2022}.
The primary goal of these efforts -- collectively called nanoscale MRI (nanoMRI) --  is to achieve {three-dimensional angstrom-resolution} imaging of macromolecules and nanoscale molecular complexes, with a keen focus on biologically significant samples, such as proteins and virus particles whose characteristic length scale is of order $(\SI{100}{\nano\meter})^3$.  In recent years, significant advancements in detection sensitivity have been made through the development of more sensitive spin sensors.
This progress ranges from {the fabrication of ultra-low} dissipation nanomechanical cantilevers \cite{Tao2015, Heritier2018, Sahafi2020} and high gradient magnets in force-detected platforms \cite{Tao2016,Mamin2012}, to improved coherence times and surface properties of shallow nitrogen-vacancy (NV) quantum sensors \cite{Sangtawesin2019, Joos2022, Zheng2022}.
While these approaches show promise, substantial progress is still needed to fully realize nanoMRI's potential as a non-destructive and chemically-selective tool for studying nanoscale nuclear spin systems at atomic length scales.

Owing to the small nuclear spin polarization, even at high field and cryogenic conditions, NMR has been plagued by low detection signal-to-noise ratio (SNR). Since its inception, numerous techniques have been devised for enhancing the  nuclear spin polarization in NMR \cite{JHLee2014}. Among these, dynamic nuclear polarization (DNP) \cite{JEills2023, LThankamony2017,TVCan2015} enhances the signal by transferring the larger polarization of unpaired electrons, in either thermally, optically, or chemically polarized states, to nearby nuclei via hyperfine couplings.
DNP can boost the detection sensitivity of nuclei by a factor of $\gamma_e/\gamma_n \geq 658$ relative to the signal obtained from thermal polarization, where $\gamma_e$ and $\gamma_n$ are the electron and nuclear gyromagnetic ratios, respectively.
Such signal enhancements have shortened acquisition times by several orders of magnitude and dramatically expanded the horizons of NMR spectroscopy and imaging \cite{Gupta2016,Lesage2010,Rossini2014,Berruyer2017,Golman2003,Krishna2002,Lee2013,Ji2017,Chen2013,Nelson2013}.

\begin{figure}[h!]
    \centering
    \includegraphics[width = 0.55\textwidth]{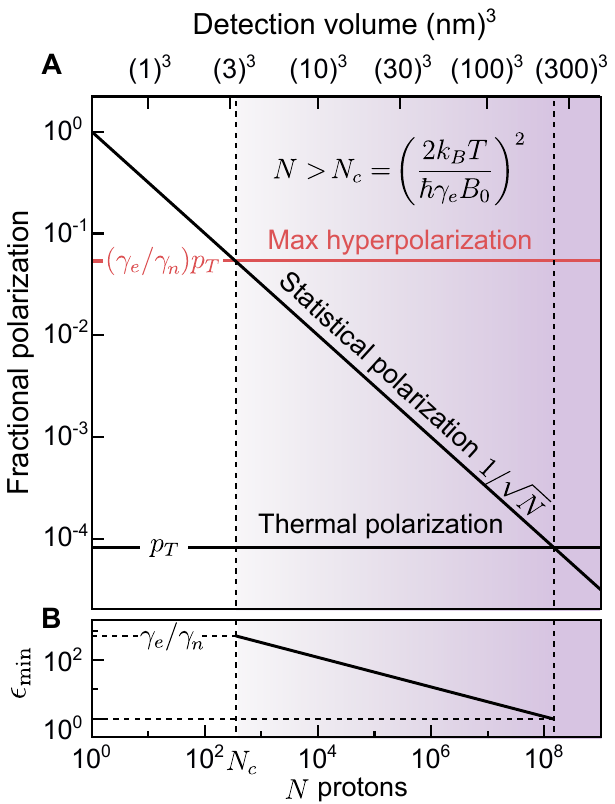}
    \caption{\textbf{Comparison of Thermal, Statistical and DNP-Enhanced Fractional Polarizations in Nanoscale Spin Ensembles.}
    (\textbf{A}) Calculated fractional polarizations for $N$ proton spins at $T = 4.2$~K and $B_0 = 0.32$~T.
    The thermal polarization $p_T = \hbar \gamma_n B_0/(2k_B T) = 7.78\times 10^{-5}$ can be enhanced by a maximum factor of $\gamma_e/\gamma_n = 658$ (red line) through polarization transfer from electrons in a thermal state.
    A potential increase in SNR via DNP can be expected for $N > N_c$ (shaded region), where $N_c = \left(\frac{2k_B T}{\hbar \gamma_e B_0}\right)^2 = 382$.
    The top axis is calculated using a representative proton density of \SI{1e10}{\m^{-3}} for organic samples, such as the tobacco mosaic virus (TMV) particle \cite{Degen2009}.
    (\textbf{B}) The minimum enhancement $\epsilon_\text{min}$ for DNP to surpass statistical polarization.
    For the example of the TMV particle occupying a $\sim \mathrm{(50 \ nm)}^3$ volume, $\epsilon_\text{min}\sim 10$.
    }
    \label{fig:HyperpolarizedVsStatisticalTheory}
\end{figure}

 The integration of DNP methods into nanoMRI marks a significant milestone towards the goal of nanoscale imaging of proteins and virus particles.
 However, achieving an SNR advantage with DNP in nanoscale systems is challenging because, as shown in Fig.~\ref{fig:HyperpolarizedVsStatisticalTheory}, for typical spin populations in nanoscale volumes,  statistical fluctuations in the  number of up vs. down spins is often much larger than the Boltzmann polarization that is commonly measured in NMR.
 Furthermore, a collection of spins is always ``statistically-polarized'', and hence can be continuously measured, whereas a measurement of the Boltzmann polarization requires waiting a time of order the spin-lattice relaxation time $T_1$ for the spin ensemble to reach thermal equilibrium. Therefore, most nanoscale force-detected NMR protocols do not measure thermal polarization, but instead rely on the detection of the statistical polarization \cite{Mamin2003,Herzog2014,Staudenmaier2023}. 
For DNP to yield an SNR advantage, it must not only boost the average polarization to a level that is larger than the statistical polarization, but also account for the higher acquisition rate of the statistically-polarized signal.

In this work, we present force-detected pulsed DNP measurements of proton spins in nanoscale droplets containing the trityl-OX063 free radical dissolved in water sugar mixtures.
We demonstrate enhancements of thermal polarization exceeding 100 at $\SI{6}{\kelvin}$ and $\SI{0.33}{\tesla}$. Importantly, we achieve a 15-fold increase in the detection SNR of hyperpolarized vs. statistically-polarized protons for a sample volume of $\sim(\SI{250}{\nano\meter})^3$, corresponding to a 200-fold decrease in the signal acquisition time.
We find that the observed enhancements are facilitated by increased spin-lattice relaxation rates of OX063 electron spins caused by spin diffusion to fast-relaxing paramagnetic defects on the surface of the silicon nanowire (SiNW) force sensor.
Numerical simulations of this process, provided in the Supplementary Materials, offer insights into optimizing the DNP efficiency in future experiments.
Our results represent a transformative step towards bringing the large sensitivity enhancements of DNP to nanoMRI measurements, paving the way for a practical NMR platform for performing three-dimensional nanoscale nuclear spin imaging.

\section*{Results}
\subsubsection*{Experimental Setup}
\begin{figure}[h]
    \centering
    \includegraphics[width = 0.75\textwidth]{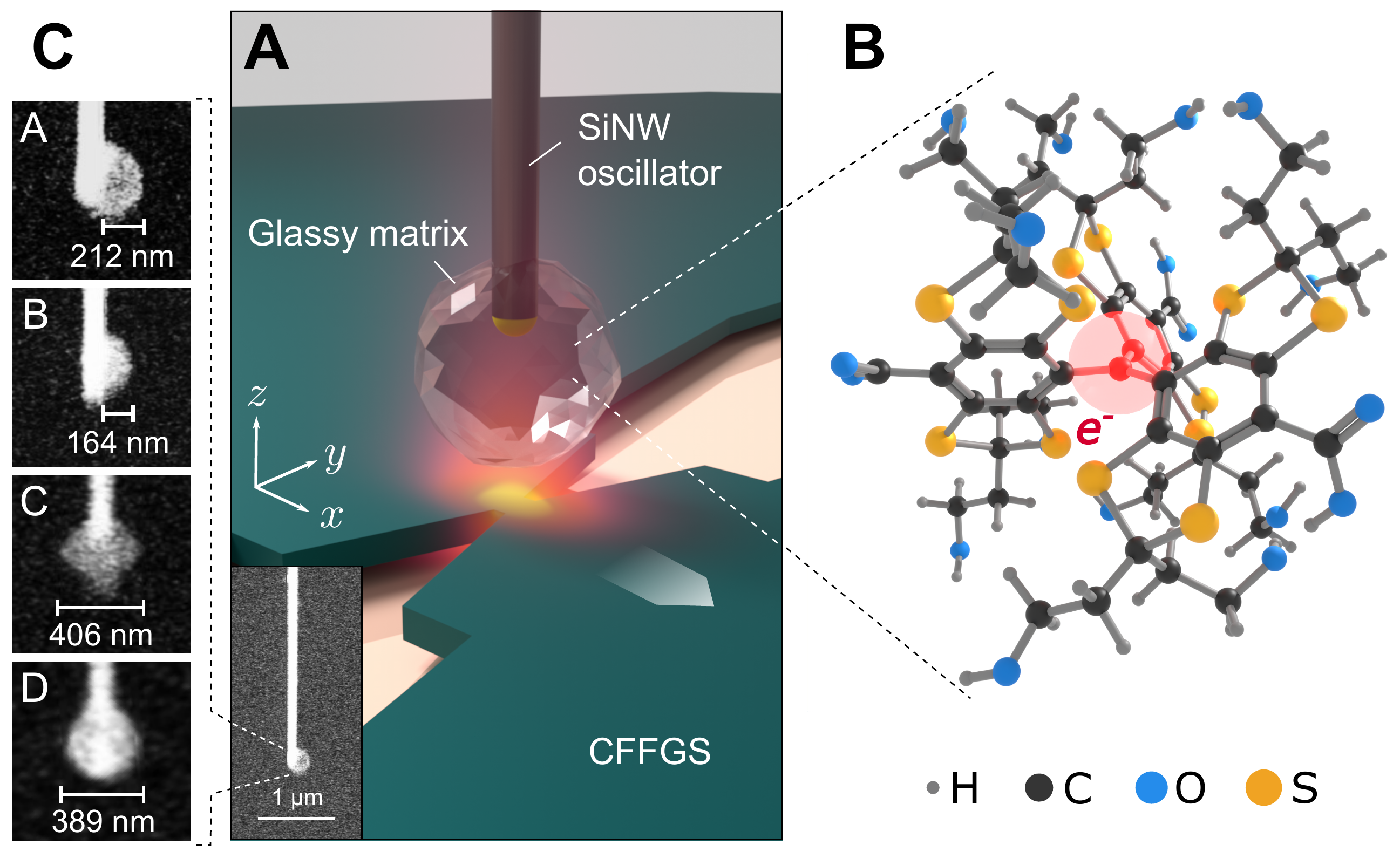}
    \caption{\textbf{Experimental Setup and Polarizing Agent.} 
    (\textbf{A}) Schematic of the experimental setup, including the SiNW force sensor, CFFGS, and the sample comprised of polarizing agents in a glassy sugar matrix. (Inset) Scanning electron microscope (SEM) image of sample~A attached to a SiNW.
    (\textbf{B}) The trityl-OX063 radical used as a polarizing agent, with the unpaired electron indicated in red.
    The radical's highly symmetric structure is known to yield an exceptionally small linewidth and g-anisotropy \cite{Lumata2013}.
    (\textbf{C}) SEM images of each of the measured nanodroplets.
    }
    \label{fig:ExpSetup}
\end{figure}

Fig.~\ref{fig:ExpSetup}A depicts our experimental setup, which is similar to the one used in our previous experiments \cite{Rose2018,Haas2021}.
Force-detected magnetic resonance measurements were performed on samples consisting of a droplet of a water/ethanol/sugar solution attached to the tip of a SiNW nanomechanical resonator. Four different samples were measured, with each droplet ranging from $\SI{150}{\nano\meter}$ to $\SI{300}{\nano\meter}$ in diameter. The samples contained varying concentrations of the trityl-OX063 electron free radical used for the DNP measurements.
The exact description of each sample is provided in Materials and Methods (Table~\ref{tab:samples}).
After attachment, the samples were mounted to the force microscope and kept under high vacuum for several days prior to cooling down to $\SI{4.2}{\kelvin}$.
Noise thermometry measurements showed an increase in the sample temperature between $\qtyrange{1}{2}{\kelvin}$ caused by absorption of $\SI{1510}{\nano\meter}$ light used for interferometric displacement detection of the SiNW -- a known effect in our platform \cite{Rose2018}.
For details regarding sample preparation and noise thermometry measurements, see the Supplementary Materials.

DNP measurements were conducted in an external field of $\SI{0.33}{\tesla}$, applied parallel to the SiNW axis.
The proton ($I=1/2$) and electron ($S=1/2$) Larmor frequencies at this field were 
$\omega_{0I}/(2\pi)=\SI{14.1}{\MHz}$ and $\omega_{0S}/(2\pi)=\SI{9.26}{\GHz}$, respectively.
A current focusing field gradient source (CFFGS) was utilized to generate the time-dependent magnetic fields and magnetic field gradients needed for spin control and detection (see Materials and Methods). The tip of the sample was aligned to the center of the CFFGS and positioned $\SI{100}{\nano\meter}$ above the surface.
The transverse field $B_1=\sqrt{B_x^2+B_y^2}/2$ used to excite magnetic resonance was generated by driving the CFFGS at the Larmor frequency of the targeted spin, where $B_x$, $B_y$, $B_z$ are the magnetic field components.
The $B_1$ field generated was highly non-uniform and varied by a factor of $\sim 2.5$ over the sample volume.
The maximum peak current applied to the CFFGS for proton and electron control was \SI{29}{\mA} and \SI{6}{\mA}, respectively.
For further details regarding the CFFGS field profile, see the Supplementary Materials.

\subsubsection*{DNP Protocol}
\begin{figure}[t]
    \centering
    \includegraphics[width =\textwidth]{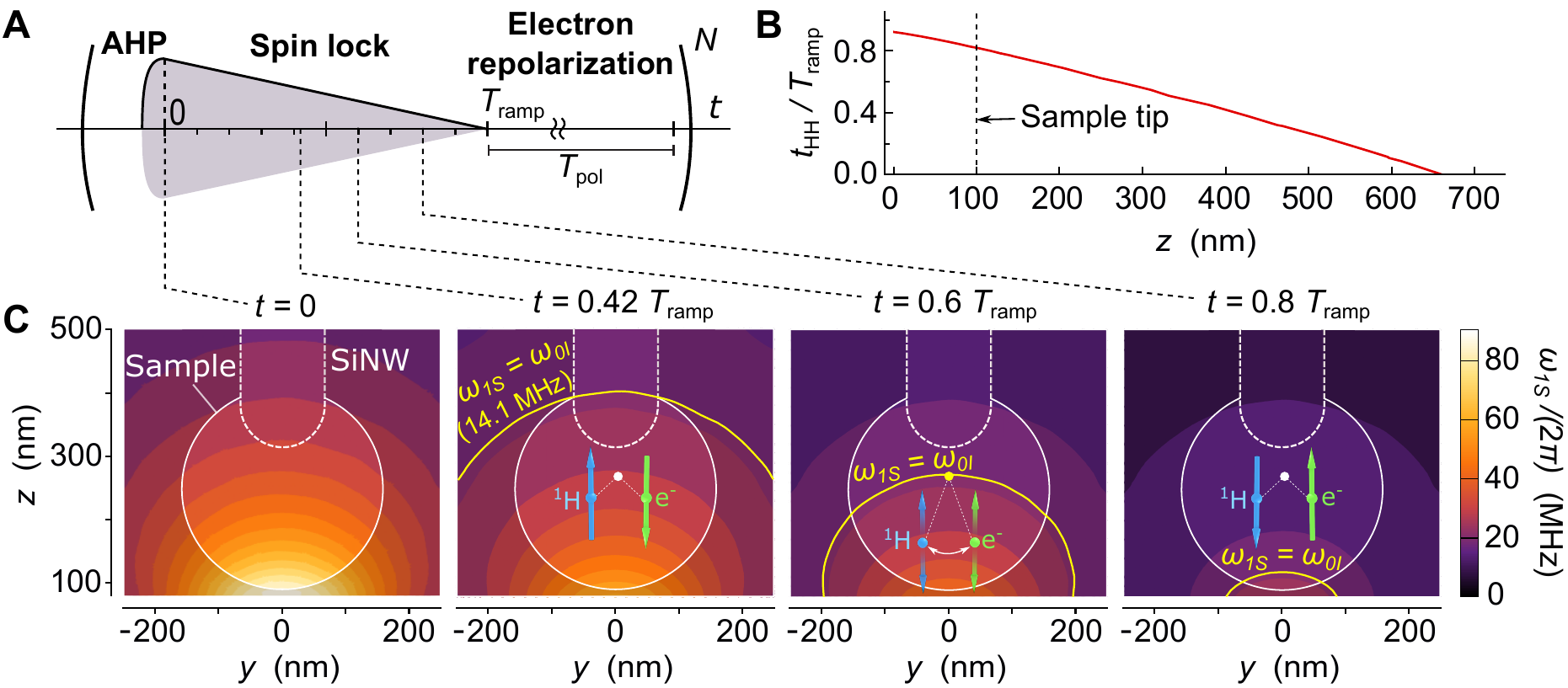}
    \caption{\textbf{RA-NOVEL Protocol.}
    (\textbf{A}) RA-NOVEL waveform applied to electrons, consisting of an adiabatic half passage (AHP), ramped-amplitude spin-locking pulse and electron repolarization period $T_\mathrm{pol}$.
    Repetitions of this primitive allow for successive build-up of proton polarization up to a time of order the proton spin-lattice relaxation time $T_{1p}$.
    (\textbf{B}) The time $t_{HH}$ at which an electron-proton pair at distance $z$ from the CFFGS undergoes the Hartmann-Hahn condition $\omega_{1S} = \omega_{0I}$. The plot is calculated along the center of the CFFGS ($x = y= 0$).
    (\textbf{C}) $\omega_{1S}$ contours at different instances in the ramp.
    An electron-proton spin pair (white dot) undergoes an adiabatic exchange of polarization as $\omega_{1S}$ is swept past the Hartmann-Hahn condition (yellow contour).
    The arrows indicate the two-spin state in the tilted rotating frame.
    The polarization transfer begins away from the CFFGS and moves towards it as the ramp progresses.
    }
    \label{fig:DNPProtocol}
\end{figure}

Our measurements utilize the nuclear orientation via electron spin-locking (NOVEL) DNP protocol---a fast and highly efficient method for polarization transfer \cite{TVCan2015,Can2017,Corzilius2020}.
For efficient polarization transfer that is robust to the spatial variation of the electron Rabi frequency $\omega_{1S} = \gamma_e B_1$ in our setup, we utilized a modified {version of the} ramped-amplitude NOVEL (RA-NOVEL) protocol \cite{Can2017}.
The protocol (Fig.~\ref{fig:DNPProtocol}A) starts by tipping the electron spins into the transverse plane with an adiabatic half passage (AHP) pulse, and applying a resonant spin-locking drive, whose amplitude is linearly swept to zero in a time $T_\mathrm{ramp}$.
The Hamiltonian of a proton-electron spin pair in the electron's rotating frame is $H(t) = - \omega_{0I} I_z - \omega_{1S}(t) S_x - (A_0 I_z + A_1 I_x) S_z,$ where $A_0$ and $A_1$ are the secular and pseudo-secular hyperfine couplings, respectively.
For protons in the OX063 molecule, we expect $A_1/(2\pi)\sim 1$~MHz \cite{Mathies2016}, while the coupling to protons in the surrounding matrix will depend on their local concentration and spatial distribution.
A rotation of the reference frame around the $y$ axis by $-\pi/2$, followed by perturbative truncation results in the polarization transfer Hamiltonian 
\begin{equation}
    \tilde{H}(t) = - \omega_{0I} I_z - \omega_{1S}(t) S_z - A_1 (I_+ S_- + I_- S_+)/2
\end{equation}
in the tilted rotating frame \cite{Can2017}, where $I_\pm= I_x \pm i I_y$ and $S_\pm = S_x \pm i S_y$ are spin ladder operators.
For an electron spin satisfying $\omega_{1S}(0) \gg \omega_{0I}$, an amplitude ramp that is sufficiently slow relative to $A_1$ induces polarization exchange as $\omega_{1S}$ passes through the Hartmann-Hahn resonance $\omega_{1S}(t) = \omega_{0I}$.
This occurs via the adiabatic evolution of the $\tilde{H}$ eigenstates from $\ket{\uparrow \downarrow}$ to $\ket{\downarrow\uparrow}$ (and vice versa), with $\ket{\downarrow \downarrow}$ and $\ket{\uparrow \uparrow}$ left invariant.
Fig.~\ref{fig:DNPProtocol}C illustrates this process at an arbitrary position in the CFFGS field distribution.
Successive build-up of the polarization can be achieved by repeated application of the RA-NOVEL waveform primitive, while including a wait time $T_\mathrm{pol}$ in between {the field sweeps} to allow for the electrons to re-polarize, and the enhanced proton polarization to diffuse through the sample.

\subsubsection*{DNP Measurements}
\begin{figure}
    \centering
    \includegraphics[width =\textwidth]{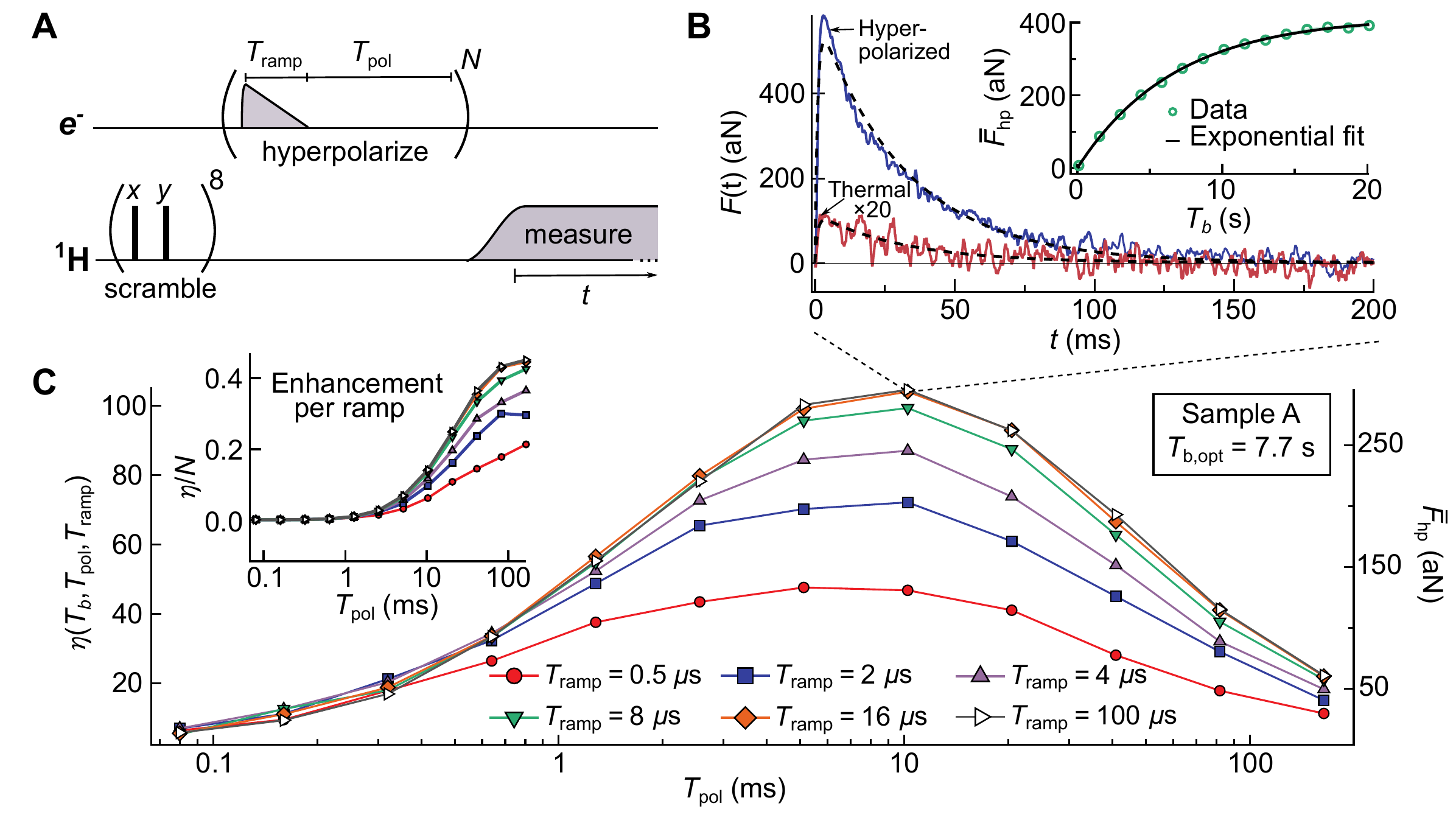}
    \caption{\textbf{Nanoscale DNP}.
    (\textbf{A}) DNP measurement sequence.
    The proton polarization is scrambled using a combination of $\pi/2)_x$ and $\pi/2)_y$ pulses with a dephasing time in between.
    DNP is then carried out by successive repetitions of the RA-NOVEL primitive for a build-up time $T_b$.
    (\textbf{B}) Measured hyperpolarized and thermal time records for sample A.
    The thermal measurement is done using the pulse sequence of (\textbf{A}) with the DNP waveform amplitude set to zero.
    A fit model to the mechanical response of the SiNW (dashed black lines) indicates a proton correlation time $\tau_m = 31.9(2)$~ms during the MAGGIC signal acquisition period (see Supplementary Materials).
    (Inset) Enhanced signal as a function of build-up time.
    (\textbf{C}) Measured enhancement as a function of $T_\text{pol}$ and $T_\text{ramp}$, for a DNP buildup time of $T_{b\text{,opt}} = \SI{7.7}{\s}$.
    (Inset) The data is processed into the enhancement per ramp, depicting saturation of the enhancement with increasing $T_\text{pol}$ and $T_\text{ramp}$.
    The error bars in the data are smaller than the plot marker size.
    }
    \label{fig:DNPMeasurements}
\end{figure}

We measured the performance of RA-NOVEL using the pulse sequence shown in Fig.~\ref{fig:DNPMeasurements}A.
After initially scrambling  the thermal proton polarization, $N$ repetitions of the RA-NOVEL waveform were applied over a build-up time $T_b = N (T_\text{ramp}+T_\text{pol})$, and the average proton polarization was measured using the modulated alternating gradients generated with currents (MAGGIC) spin detection protocol.
We refer to the Supplementary Materials and Ref.~\cite{Haas2021,Rose2018} for details regarding the MAGGIC protocol.
The MAGGIC protocol couples the average longitudinal ($z$ axis) spin polarization to {the mechanical motion of the SiNW via the time-dependent} force $F(t) \propto e^{-t/\tau_m} \int d^3 r~ n(\mathbf{r}) G(\mathbf{r}) m_0(\mathbf{r})$, resonant with the SiNW force sensor.
Here, $G(\mathbf{r}) = \partial B_z(\mathbf{r})/\partial y$ is the peak amplitude of the detection gradient, $n(\mathbf{r})$ is the spin density, $\tau_m$ is the spin correlation time under MAGGIC, and $m_0(\mathbf{r})$ is the average polarization of a spin at position $\mathbf{r}$ at the start of the measurement.
We construct the average measured force $\bar{F} = \int_0^{\tau_0} dt~w(t)F(t)$ for a given time record $F(t)$, where $\tau_0$ is the measurement duration and $w(t)$ is a matched filter kernel determined for optimal SNR (see Supplementary Materials).

We quantify the {DNP} enhancement as the ratio $\eta(T_b, T_\text{pol},T_\text{ramp}) = \bar{F}_\text{hp}(T_b, T_\text{pol},T_\text{ramp})/\bar{F}_\text{th}(T_b)$, where we compare the hyperpolarized signal $\bar{F}_\text{hp}(T_b, T_\text{pol},T_\text{ramp})$ with the thermal signal $\bar{F}_\text{th}(T_b)$ acquired from thermally-polarized protons for the same duration $T_b$.
In the main text, we focus on the data for sample A, while providing a summary for all samples in Table~\ref{tab:dataSummary}, and referring to the Supplementary Materials for complete data sets.
Fig.~\ref{fig:DNPMeasurements}B shows the measured  time records $F(t)$ with and without DNP.
The hyperpolarized and thermal signals were acquired with \SI{8}{\min} and \SI{17.5}{\hour} of averaging, respectively.
The optimum DNP buildup time $T_{b\text{,opt}}$ was determined by measuring the time constant $\tau_{\text{DNP}}$ of the DNP buildup (Fig.~\ref{fig:DNPMeasurements}A(inset)) and choosing the value of $T_b$ that maximized the detection $\text{SNR}\propto\sqrt{N_\text{avg}}(1-e^{-T_b/\tau_{\text{DNP}}})$.
Here, $N_\text{avg} = T_e/T_b$ is the number of averages during the total experiment duration $T_e$.
For the measurements shown in Fig.~\ref{fig:DNPMeasurements}, $\tau_\text{DNP}= \SI{6.7 \pm 0.1}{\s}$ and $T_{b\text{,opt}} = \SI{7.7}{\s}$.
The optimum DNP parameters were found by measuring $\eta$ for different values of $T_{\text{pol}}$ and $T_{\text{ramp}}$ with $T_b = T_{b \text{,opt}}$ held constant.
The measured data (Fig.~\ref{fig:DNPMeasurements}C) shows two qualitative features of the DNP process. First, the maximum value of $\eta$ increases with increasing $T_{\text{ramp}}$, until either $T_{\text{ramp}}$ exceeds the time scale set by the hyperfine coupling, or it approaches the electron spin-lattice relaxation time in the rotating frame $T_{1\rho e}$. Second, there is a trade-off between allowing enough time for electron re-polarization (governed by the electron spin-lattice relaxation time $T_{1e}$) and maximizing the number of DNP repetitions within the available buildup time $T_b$.
For this sample, with $T_{1e} = \SI{20.5 \pm 0.3}{\ms}$, a maximum enhancement of $\eta_\text{opt} = \SI{105 \pm 8}{}$ was achieved for $T_{\text{pol,opt}} = \SI{10.24}{\ms}$ and $T_{\text{ramp,opt}} =\SI{16}{\micro\s}$.
The ramp time \(T_{\mathrm{ramp,opt}}\) was 1-2 orders of magnitude lower than $T_{1\rho e}$ for all samples, indicating that \(T_{\mathrm{ramp,opt}}\) was set by the timescale of the pseudo-secular hyperfine coupling.
We provide a semiclassical model for the DNP enhancement in the Supplementary Materials.

Previous force-detected DNP measurements of proton spins performed at $\SI{4.2}{\kelvin}$ and $\SI{0.6}{\tesla}$ have obtained enhancements between 10-20 \cite{Issac2016}.
The DNP efficiency in these measurements may have been limited by the significant magnetic field inhomogeneity generated by the permanent magnets utilized in these studies, potentially hindering spin diffusion among proton spins. A distinguishing feature of our measurement platform is the use of the CFFGS, which permits full temporal control of the detection gradient and the control fields applied to the spins.
The ability to turn off the large static field gradient during the DNP process may have contributed to the larger enhancements obtained in our measurements.

\begin{table}[h!]
\centering
\caption{\textbf{Summary of nanoscale DNP results}, including measured DNP performance metrics and spin relaxation times.
See the Supplementary Materials for details of the relaxation time measurements.}

\vspace{10pt}
{\small 
\begin{tabular}{l*5l}
\toprule
 & \textbf{Sample} & \textbf{A} & \textbf{B} & \textbf{C} & \textbf{D}\\
\midrule
\multirow{5}{*}{DNP performance} 
& $\eta_\text{opt}$ & $105(8)$ & $118(8)$ & $78(3)$ & $40.6(4)$\\
& $\tau_\text{DNP}$~(s) & $6.7(1)$ & $0.67(1)$ & $8.4(2)$ & $14.9(5)$\\
& $T_{b\text{,opt}}$~(s) & $7.7$ & $0.8$ & $10.5$ & $20.8$\\
& $T_\text{pol,opt}$~(ms) & $10.24$ & $1.28$ & $10.24$ & $5.12$\\
& $T_\text{ramp,tabopt}$~(\SI{}{\micro\s}) & $16$ & $8$ & $16$ & $100$\\
\midrule
\multirow{3}{*}{Relaxation times} 
& $T_{1p}$~(s) & $10.2(3)$ & $1.01(3)$ & $10.8(6)$ & $20(1)$\\
& $T_{1\rho e}$~(ms) & $2.5(1)$ & $0.45(4)$ & $3.5(1)$ & $7.5(3)$\\
& $T_{1e}$~(ms) & $20.5(3)$ & $2.5(1)$ & $28.6(7)$ & $30.0(4)$\\
\bottomrule
\end{tabular}
}
\label{tab:dataSummary}
\end{table}

\subsubsection*{Comparison of DNP-Enhanced and Statistical Polarization SNR}
\begin{figure}[H]
    \centering
    \includegraphics[width =\textwidth]{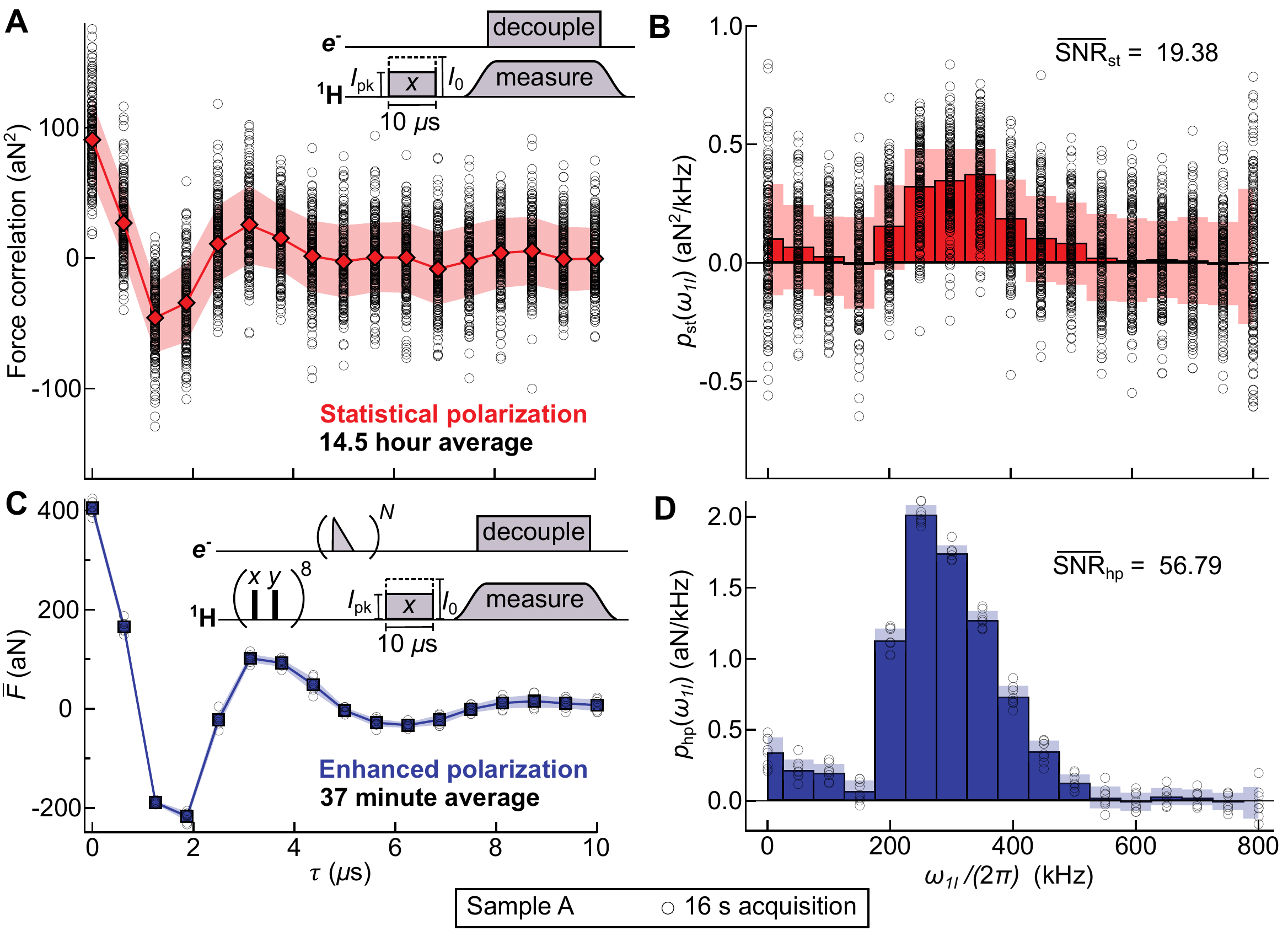}
    \caption{\textbf{DNP-Enhanced and Statistical Polarization SNR Comparison.}
    Measured time-domain Fourier encoding data for (\textbf{A}) statistically-polarized and (\textbf{C}) hyperpolarized  proton spins in sample A. The corresponding Rabi frequency distributions are shown in (\textbf{B}) and (\textbf{D}). The frequency resolution of each distribution is \SI{50}{\kHz}.
    The shaded regions indicate the standard deviation of the acquired data (open circles) for each point in the time record, and the corresponding Rabi frequency distributions. 
    The insets show the pulse sequence used for the two types of measurement.
    A constant amplitude excitation was applied at $\omega_{0S}$ to decouple the electrons while acquiring the proton signal. The decoupling procedure resulted in extending the proton spin correlation time from \qtyrange{31.9}{135.6}{\ms} during the acquisition period.
    }
    \label{fig:nanoMRI}
\end{figure}

To quantify the improvement in detection sensitivity provided by DNP relative to statistical polarization, we measured the proton Rabi distribution by using the $B_1$ gradient generated by the CFFGS to phase encode the proton spins. 
Following the Fourier-encoding scheme presented in Ref.~\cite{Rose2018,Haas2021}, time-domain signals were acquired, with and without prior hyperpolarization, after applying \SI{10}{\micro\s} long resonant pulses to the proton spins. The amplitude $I_\text{pk}$ of the encoding pulses were varied between \qtyrange{0}{29}{\mA} in 17 steps and the spin signal was recorded as a function of the effective encoding time $\tau = \SI{10}{\micro\s} \times I_\text{pk}/I_0$, with $I_0 = \SI{29}{mA}$. Fig.~\ref{fig:nanoMRI} shows the SNR comparison for sample A. Prior to the acquisition of the DNP data, protons were hyperpolarized for \SI{7.7}{\s} by applying 752 repetitions of the DNP waveform, with $T_{\text{ramp}}=\SI{16}{\micro\s}$ and $T_{\text{pol}}=\SI{10.24}{\ms}$.
After hyperpolarization, the proton signal was acquired for \SI{400}{\ms}. Each data point, shown by the open circles in Fig.~\ref{fig:nanoMRI}C, represents the average of two such measurements, corresponding to a total measurement time of \SI{16.3}{\s}. The statistically-polarized data was obtained by continuously acquiring the spin signal for \SI{16.1}{\s}, calculating the average force in $100$~ms long blocks of data acquired before and after the application of the \SI{10}{\micro\s} long resonant pulse, and calculating the average correlation of two consecutive measurements; the average correlation is indicated by the open circles in Fig.~\ref{fig:nanoMRI}A.

The solid lines in Fig.~\ref{fig:nanoMRI}A,C represent the average of 192 (statistical) and 8 (hyperpolarized) data points measured for each value of $\tau$. The averages shown correspond to a total measurement time of \SI{14.5}{\hour} (statistical) and \SI{37}{\min} (hyperpolarized). 
Denoting the measured Rabi-frequency distribution and its standard deviation by $p(\omega_{1I})$ and $\sigma_{p}(\omega_{1I})$, respectively, we define the average  SNR as
 $\mathrm{\overline{SNR}} = \sum_{i = 1}^{17} p(\omega_i)~\mathrm{SNR}_i /  \sum_{i = 1}^{17} p(\omega_i)$, where $\omega_i$ is the $i^{\text{th}}$ Rabi frequency bin and $\mathrm{SNR}_i = p(\omega_i)/\sigma_{p}(\omega_i)$.
We determine $\mathrm{\overline{SNR}_{st}} = 19.38$ and $\mathrm{\overline{SNR}_{hp}} = 56.79$ for the statistical and hyperpolarized data, respectively, revealing a reduction in averaging time by a factor of $({\mathrm{\overline{SNR}_{hp}}}/{\mathrm{\overline{SNR}_{st}}})^2 \times (14.5~\text{h}/37~\text{min}) = 204$ to achieve equal average SNRs.
Identical measurements on sample C, presented in the Supplementary Materials, indicate a similar 189-fold reduction in averaging time.
A theoretical comparison of SNR for statistical and hyperpolarized measurements is provided in the Supplementary Materials.

\subsubsection*{Surface-Induced Electron Spin Relaxation}

The electron spin-lattice relaxation time $T_{1e}$  plays an important role in the DNP process, as it sets an upper bound on the repetition rate of the DNP waveform. A previous study conducted with bulk OX063 samples at \SI{6}{\kelvin} for concentrations between \qtyrange{1}{60}{\milli\Molar} observed $T_{1e}$ values ranging from $\SI{7}{\s}$ to $\SI{50}{\ms}$ \cite{Chen2016}. The primary relaxation mechanism in these studies was attributed to cross-relaxation with a small number of fast-relaxing OX063 aggregates in the solution. We find the $T_{1e}$ values measured in nanoscale sugar droplets at the same temperature and over a similar concentration range of \qtyrange{1.3}{49.4}{\milli\Molar} to be substantially shorter than these previous studies, varying between \qtyrange{70}{2.5}{\milli\s}. Moreover, we find that the value of $T_{1e}$ for samples prepared from the same solution on two different SiNWs can vary by more than a factor of 10.

In investigating the possible sources that could account for the shorter relaxation times, we first measured the thermal electron spin polarization as a function of the SiNW temperature. For these studies, we used a \SI{1.3}{\milli\Molar} concentration reference sample, which had the longest relaxation time $(T_{1e}=\SI{70}{\milli\s})$ of all the samples we measured, and therefore should exhibit the highest sensitivity to extrinsic relaxation mechanisms. As seen in Fig.~S7B of the Supplementary Materials, the thermal polarization varies inversely with the temperature of the SiNW, consistent with the Curie-Weiss Law, indicating that the additional relaxation mechanism is in thermal equilibrium with the SiNW. This observation rules out non-thermal sources of noise, e.g., excess noise from the microwave amplifiers, as the cause for the enhanced $T_{1e}$ relaxation.

Previous studies conducted with nitrogen-vacancy (NV) centers have shown that Johnson noise caused by the thermal motion of electrons in a metal can lead to faster spin-lattice relaxation rates \cite{Kolkowitz2015}.
To see whether the proximity to the CFFGS had an effect, we measured $T_{1e}$ as a function of the distance from the CFFGS. Our measurements (Fig.~S8 of the Supplementary Materials) show no variation over a large range of tip-surface distances between \qtyrange{50}{500}{\nano\meter}, thus ruling out this effect.
For details of the $T_{1e}$ vs. distance and spin thermometry measurements, see the Supplementary Materials.

Finally, it is known that the $\mathrm{Si/SiO_2}$ interface contains fast-relaxing paramagnetic defects (dangling bonds) \cite{Wang2006, Baumer2004}, where the defect surface density $\sigma_d$ can vary from \SI{e12}{\per\square\centi\meter} to \SI{e13}{\per\square\centi\meter} \cite{Brower1986,Nouwen2000,Rurali2010,Stesmans1989}, and where the defect spin-lattice relaxation time has been measured to be of order $T_{1d} \approx \SI{40}{\micro\s}$ at \SI{4}{\kelvin} \cite{Stesmans1989}.
To investigate the role played by paramagnetic defects, we conducted semi-classical simulations of spin diffusion, wherein electron spins on the surface of the SiNW and the OX063 were evolved probabilistically according to their dipole-dipole interactions \cite{Bloembergen1949}, accounting for the sample geometry, electron spin lineshape, and surface defect density. 
We find that for simulations that assume $T_{1d}=\SI{50}{\micro\s}$ and $\sigma_d = \SI{1e13}{\per\square\centi\metre}$, the calculated and measured $T_{1e}$ values for the OX063 spins are consistent with each other (see Supplementary Materials Section~11). Interestingly, the simulations also reveal that the average relaxation rate of the OX063 spins can be very sensitive to the shape of the droplet and its placement on the SiNW, consistent with our observations. Simulations conducted for samples B and C, which were prepared using the same stock solution, reproduce the factor of 10 variation in $T_{1e}$ observed in the measurements.
The variation is attributed to sample~B being smaller and having a larger contact area with the SiNW, whereas sample~C is somewhat larger, with the SiNW having minimal entry into the droplet (Fig.~\ref{fig:ExpSetup}C).
Our simulations point to cross relaxation among paramagnetic defects on the surface of the SiNW and the OX063 spins as the dominant source for the reduction in $T_{1e}$ observed in our measurements.

\section*{Discussion}

We have demonstrated large sensitivity enhancements in proton spin measurements by integrating high efficiency pulsed DNP into a force-detected nanoMRI platform. Utilizing this enhanced polarization, we have achieved a 204-fold reduction in acquisition time in measuring the proton Rabi frequency distribution relative to measurements using statistical polarization.
Our findings suggest that the enhanced spin relaxation caused by defect spins on the SiNW surface significantly reduces the $T_{1e}$ of the OX063 electron spins at \SI{6}{\kelvin} and \SI{0.33}{\tesla}, contributing to the large DNP enhancements observed in this work. Previous DNP measurements conducted at \SI{1.4}{\kelvin} and \SI{2.35}{\tesla} with silicon microparticles\cite{Dementyev2008} 
have demonstrated that spin diffusion to a silicon surface, and subsequent relaxation via paramagnetic defects can lead to an increase in DNP enhancement. In a similar manner,
it should be possible to further exploit the rapid spin relaxation caused by proximity to the SiNW surface to maintain large DNP enhancement at lower temperatures and higher magnetic fields. 

In addition to the large sensitivity enhancement, our results are also notable as a demonstration of the versatility of our nanoMRI setup.  
We demonstrate that it is possible to control and measure both electron and proton spins in the same setup -- a significant advancement over previous work \cite{Rose2018} where measurements were performed only on proton spins. Furthermore, we demonstrate the ability to reproducibly perform sample attachment with nanoscale droplets on the SiNWs. We envision that similar methods can be used to load other biologically relevant molecules e.g., proteins or virus particles, to our nanoMRI setup.

While these results already mark a transformative capability in nanoMRI detection, there remains significant room for improvement.
Pulsed DNP schemes such as RA-NOVEL are known for their efficiency at high magnetic fields \cite{JEills2023,Mathies2016}.
Extending our experiments to higher fields and sub-Kelvin temperatures could usher a fundamentally new regime for nuclear spin detection sensitivity.
For example, polarization transfer with the same efficiency at \SI{1}{\tesla} and $0.3$~K would lead to a factor of 400 SNR improvement compared to the detection of statistical polarization --- a pivotal advance that would unlock unprecedented opportunities for probing nanoscale nuclear spin systems with near-atomic resolution.

\section*{Materials and Methods}

\subsubsection*{SiNW Resonator}

SiNW arrays were epitaxially grown on a Si(111) substrate with dimensions \(1.5~\text{mm}\times1~\text{mm}\times0.4~\text{mm}\), using the vapor-liquid-solid method with a patterned array of Au catalyst particles near the edge of the chip.
The growth environment consisted of H\(_2\) and HCl gases with a SiH\(_4\) precursor.
The resulting SiNWs were \SI{18}{\micro\m} long with a diameter of 120~nm \cite{Sahafi2020}.
The Au catalyst was removed using a KI solution to prevent the interaction of the Au particle with the electromagnetic fields generated by the CFFGS. 
Before sample attachment, the frequency of the fundamental flexural mode of the SiNWs ranged from 300~kHz to 600~kHz.
The frequency decreased by 50-100~kHz after attachment of the sugar droplets, due to mass loading.
After sample attachment, the quality factor of the SiNWs was $\sim 30,000$ at \SI{4}{\kelvin}.
To increase the measurement bandwidth of the resonator, a feedback signal was applied to a piezoelectric transducer glued to the SiNW chip, which effectively reduced the quality factor to $900$ \cite{Poggio2007}.

\subsubsection*{Sample Preparation and Attachment}

The stock solutions used for sample preparation are given in Table~\ref{tab:samples}.
Droplets of the solution were attached to the SiNWs under an optical microscope using a micro-manipulator setup with a pneumatic injector.
For details of the attachment process, see the Supplementary Materials.
Scanning electron microscope images of the samples are depicted in Fig.~\ref{fig:ExpSetup}C.
Due to the uncertainty in water retention within the nanodroplet at 4~K and high-vacuum, exact calculation of the final OX063 concentrations is unfeasible.
Accordingly, a concentration range is provided in Table~\ref{tab:concentrations}, with the water content ranging from 0 to 100\% of the initial solution. 
Ethanol was added to the water sugar mixture to reduce the surface tension and facilitate penetration of the SiNW into the solution.
The ethanol is assumed to have fully evaporated under experimental conditions.

\begin{table}[h!]
\centering
\caption{\textbf{Stock solutions used for sample preparation.}}\label{tab:samples}
\begin{tabular}{*5l}
\toprule
\textbf{Sample} & \textbf{Stock Solution}\\
\midrule
A  & 0.9~M \(\mathrm{d}_{12}\)-glucose/0.1~M glucose/2.1~mM OX063 in \(\mathrm{H}_2\mathrm{O}\)/\(\mathrm{D}_2\mathrm{O}\)/\(\mathrm{d}_6\)-ethanol (1/9/10, v/v/v)\\
B, C  & 0.9~M \(\mathrm{d}_{12}\)-glucose/0.1~M glucose/5.9~mM OX063 in \(\mathrm{H}_2\mathrm{O}\)/\(\mathrm{D}_2\mathrm{O}\)/\(\mathrm{d}_6\)-ethanol (1/9/10, v/v/v)\\
D  &  1.2~M sucrose/7.3~mM OX063 in H$_2$O/ethanol (2/1, v/v)\\
\bottomrule
\end{tabular}
\end{table}

\begin{table}[h!]
\centering
\caption{\textbf{Calculated OX063 Concentrations.} Maximum and minimum values correspond to 0\% and 100\% water retention in the nanodroplets, respectively.
}
\vspace{10pt}
{\small 
\begin{tabular}{l*5l}
\toprule
 & \textbf{Sample} & \textbf{A} & \textbf{B} & \textbf{C} & \textbf{D}\\
\midrule
\multirow{2}{*}{OX063 concentration} 
& $\mathrm{min~(mM)}$ & $3.5$ & $9.7$ & $9.7$ & $9.8$\\
& $\mathrm{max~(mM)}$ & $17.8$ & $49.4$ & $49.4$ & $29.0$\\
\bottomrule
\end{tabular}
} 
\label{tab:concentrations}
\end{table}

\subsubsection*{CFFGS Device}
The CFFGS device was fabricated by electron-beam lithography and reactive ion-beam etching of a 100-nm thick Al film deposited on a sapphire substrate. The device contained a 150-nm wide and 50-nm-long constriction, which served to focus electrical currents to produce the magnetic fields used for spin detection and control. 

\subsubsection*{NMR and ESR Electronics}
We used two arbitrary waveform generators (AWGs) to generate the spin control and detection waveforms applied to the CFFGS.
One AWG was used to generate the MAGGIC waveforms at the SiNW frequency for spin detection, while the other generated NMR and ESR waveforms on separate channels.
The ESR pulses were mixed in software with a \SI{374}{\MHz} carrier and then up-converted to \SI{9.26}{\GHz} using an IQ mixer.
The signals were combined at the final stage using a diplexer and bias tees.
To avoid spurious electron relaxation, we minimized electronic noise near $\SI{9.26}{\GHz}$ by adding a fast switch to the output of the microwave amplifier, which provided a connection to the CFFGS only during application of the microwave pulses.
A schematic of the electronics is provided in the Supplementary Materials.

\subsubsection*{Force-Detected Magnetic Resonance Microscope}
Experiments were conducted under high vacuum using a custom-built force microscope operating in high vacuum at \SI{4.2}{\kelvin}.
Relative positioning between the SiNW chip, CFFGS and the optical fiber used for detecting the interferometric displacement of the SiNW was performed using a combination of nanopositioners and piezoelectric scanners.


\begin{thebibliography}{10}

\bibitem{Degen2009}
C.~L. Degen, M.~Poggio, H.~J. Mamin, C.~T. Rettner, D.~Rugar, \textit{Nanoscale
  Magnetic Resonance Imaging}.
\newblock {\it Proc. Natl. Acad. Sci. U.S.A\/} {\bf 106}, 1313 (2009).

\bibitem{Nichol2012}
J.~M. Nichol, E.~R. Hemesath, L.~J. Lauhon, R.~Budakian, \textit{Nanomechanical
  Detection of Nuclear Magnetic Resonance using a Silicon Nanowire Oscillator}.
\newblock {\it Phys. Rev. B\/} {\bf 85}, 054414 (2012).

\bibitem{Maletinsky2012}
P.~Maletinsky, S.~Hong, M.~S. Grinolds, B.~Hausmann, M.~D. Lukin, R.~L.
  Walsworth, M.~Loncar, A.~Yacoby, \textit{A Robust Scanning Diamond Sensor for
  Nanoscale Imaging with Single Nitrogen-Vacancy Centres}.
\newblock {\it Nat. Nanotechnol.\/} {\bf 7}, 320-324 (2012).

\bibitem{Nichol2013}
J.~M. Nichol, T.~R. Naibert, E.~R. Hemesath, L.~J. Lauhon, R.~Budakian,
  \textit{Nanoscale Fourier-Transform Magnetic Resonance Imaging}.
\newblock {\it Phys. Rev. X\/} {\bf 3}, 031016 (2013).

\bibitem{Haberle2015}
T.~H{\"a}berle, D.~Schmid-Lorch, F.~Reinhard, J.~Wrachtrup, \textit{Nanoscale
  Nuclear Magnetic Imaging with Chemical Contrast}.
\newblock {\it Nat. Nanotechnol.\/} {\bf 10}, 125-128 (2015).

\bibitem{Kost2015}
M.~Kost, J.~Cai, M.~B. Plenio, \textit{Resolving Single Molecule Structures
  with Nitrogen-Vacancy Centers in Diamond}.
\newblock {\it Sci. Rep.\/} {\bf 5}, 11007 (2015).

\bibitem{Ajoy2015}
A.~Ajoy, U.~Bissbort, M.~D. Lukin, R.~L. Walsworth, P.~Cappellaro,
  \textit{Atomic-Scale Nuclear Spin Imaging Using Quantum-Assisted Sensors in
  Diamond}.
\newblock {\it Phys. Rev. X\/} {\bf 5}, 011001 (2015).

\bibitem{Rugar2015}
D.~Rugar, H.~J. Mamin, M.~H. Sherwood, M.~Kim, C.~T. Rettner, K.~Ohno, D.~D.
  Awschalom, \textit{Proton Magnetic Resonance Imaging using a
  Nitrogen--Vacancy Spin Sensor}.
\newblock {\it Nat. Nanotechnol.\/} {\bf 10}, 120-124 (2015).

\bibitem{Wang2019}
P.~Wang, S.~Chen, M.~Guo, S.~Peng, M.~Wang, M.~Chen, W.~Ma, R.~Zhang, J.~Su,
  X.~Rong, F.~Shi, T.~Xu, J.~Du, \textit{Nanoscale Magnetic Imaging of
  Ferritins in a Single Cell}.
\newblock {\it Sci. Adv.\/} {\bf 5}, eaau8038 (2019).

\bibitem{Rose2018}
W.~Rose, H.~Haas, A.~Q. Chen, N.~Jeon, L.~J. Lauhon, D.~G. Cory, R.~Budakian,
  \textit{High-Resolution Nanoscale Solid-State Nuclear Magnetic Resonance
  Spectroscopy}.
\newblock {\it Phys. Rev. X\/} {\bf 8}, 011030 (2018).

\bibitem{Abobeih2019}
M.~H. Abobeih, J.~Randall, C.~E. Bradley, H.~P. Bartling, M.~A. Bakker, M.~J.
  Degen, M.~Markham, D.~J. Twitchen, T.~H. Taminiau, \textit{Atomic-Scale
  Imaging of a 27-Nuclear-Spin Cluster using a Quantum Sensor}.
\newblock {\it Nature\/} {\bf 576}, 411-415 (2019).

\bibitem{Grob2019}
U.~Grob, M.~D. Krass, M.~H{\'e}ritier, R.~Pachlatko, J.~Rhensius,
  J.~Ko{\v{s}}ata, B.~A. Moores, H.~Takahashi, A.~Eichler, C.~L. Degen,
  \textit{Magnetic Resonance Force Microscopy with a One-Dimensional Resolution
  of 0.9 Nanometers}.
\newblock {\it Nano Letters\/} {\bf 19}, 7935-7940 (2019).

\bibitem{Haas2021}
H.~Haas, S.~Tabatabaei, W.~Rose, P.~Sahafi, M.~Piscitelli, A.~Jordan,
  P.~Priyadarsi, N.~Singh, B.~Yager, P.~J. Poole, D.~Dalacu, R.~Budakian,
  \textit{Nuclear Magnetic Resonance Diffraction with Subangstrom Precision}.
\newblock {\it Proc. Natl. Acad. Sci. U.S.A\/} {\bf 119}, e2209213119 (2022).

\bibitem{Krass2022}
M.~Krass, N.~Prumbaum, R.~Pachlatko, U.~Grob, H.~Takahashi, Y.~Yamauchi, C.~L.
  Degen, A.~Eichler, \textit{Force-Detected Magnetic Resonance Imaging of
  Influenza Viruses in the Overcoupled Sensor Regime}.
\newblock {\it Phys. Rev. Appl.\/} {\bf 18}, 034052 (2022).

\bibitem{Tao2015}
Y.~Tao, C.~L. Degen, \textit{Single-Crystal Diamond Nanowire Tips for
  Ultrasensitive Force Microscopy}.
\newblock {\it Nano Lett.\/} {\bf 15}, 7893-7897 (2015).

\bibitem{Heritier2018}
M.~H{\'e}ritier, A.~Eichler, Y.~Pan, U.~Grob, I.~Shorubalko, M.~D. Krass,
  Y.~Tao, C.~L. Degen, \textit{Nanoladder Cantilevers Made from Diamond and
  Silicon}.
\newblock {\it Nano Lett.\/} {\bf 18}, 1814-1818 (2018).

\bibitem{Sahafi2020}
P.~Sahafi, W.~Rose, A.~Jordan, B.~Yager, M.~Piscitelli, R.~Budakian,
  \textit{Ultralow Dissipation Patterned Silicon Nanowire Arrays for Scanning
  Probe Microscopy}.
\newblock {\it Nano Lett.\/} {\bf 20}, 218-223 (2020).

\bibitem{Tao2016}
Y.~Tao, A.~Eichler, T.~Holzherr, C.~L. Degen, \textit{Ultrasensitive Mechanical
  Detection of Magnetic Moment using a Commercial Disk Drive Write Head}.
\newblock {\it Nat. Commun.\/} {\bf 7}, 12714 (2016).

\bibitem{Mamin2012}
H.~J. Mamin, C.~T. Rettner, M.~H. Sherwood, L.~Gao, D.~Rugar, \textit{High
  Field-Gradient Dysprosium Tips for Magnetic Resonance Force Microscopy}.
\newblock {\it Appl. Phys. Lett.\/} {\bf 100}, 013102 (2012).

\bibitem{Sangtawesin2019}
S.~Sangtawesin, B.~L. Dwyer, S.~Srinivasan, J.~J. Allred, L.~V.~H. Rodgers,
  K.~De~Greve, A.~Stacey, N.~Dontschuk, K.~M. O'Donnell, D.~Hu, D.~A. Evans,
  C.~Jaye, D.~A. Fischer, M.~L. Markham, D.~J. Twitchen, H.~Park, M.~D. Lukin,
  N.~P. de~Leon, \textit{Origins of Diamond Surface Noise Probed by Correlating
  Single-Spin Measurements with Surface Spectroscopy}.
\newblock {\it Phys. Rev. X\/} {\bf 9}, 031052 (2019).

\bibitem{Joos2022}
M.~Joos, D.~Bluvstein, Y.~Lyu, D.~Weld, A.~Bleszynski~Jayich,
  \textit{Protecting Qubit Coherence by Spectrally Engineered Driving of the
  Spin Environment}.
\newblock {\it npj Quantum Inf.\/} {\bf 8}, 47 (2022).

\bibitem{Zheng2022}
W.~Zheng, K.~Bian, X.~Chen, Y.~Shen, S.~Zhang, R.~St{\"o}hr, A.~Denisenko,
  J.~Wrachtrup, S.~Yang, Y.~Jiang, \textit{Coherence Enhancement of Solid-State
  Qubits by Local Manipulation of the Electron Spin Bath}.
\newblock {\it Nat. Phys.\/} {\bf 18}, 1317-1323 (2022).

\bibitem{JHLee2014}
J.~H. Lee, Y.~Okuno, S.~Cavagnero, \textit{Sensitivity Enhancement in Solution
  NMR: Emerging Ideas and New Frontiers}.
\newblock {\it J. Magn. Reson.\/} {\bf 241}, 18-31 (2014).

\bibitem{JEills2023}
J.~Eills, D.~Budker, S.~Cavagnero, E.~Y. Chekmenev, S.~J. Elliott, S.~Jannin,
  A.~Lesage, J.~Matysik, T.~Meersmann, T.~Prisner, J.~A. Reimer, H.~Yang, I.~V.
  Koptyug, \textit{Spin Hyperpolarization in Modern Magnetic Resonance}.
\newblock {\it Chem. Rev.\/} {\bf 123}, 1417-1551 (2023).

\bibitem{LThankamony2017}
A.~S. {Lilly Thankamony}, J.~J. Wittmann, M.~Kaushik, B.~Corzilius,
  \textit{Dynamic Nuclear Polarization for Sensitivity Enhancement in Modern
  Solid-State NMR}.
\newblock {\it Prog. Nucl. Magn. Reson. Spectrosc.\/} {\bf 102-103}, 120-195
  (2017).

\bibitem{TVCan2015}
T.~Can, Q.~Ni, R.~Griffin, \textit{Mechanisms of Dynamic Nuclear Polarization
  in Insulating Solids}.
\newblock {\it J. Magn. Reson.\/} {\bf 253}, 23-35 (2015).

\bibitem{Gupta2016}
R.~Gupta, M.~Lu, G.~Hou, M.~A. Caporini, M.~Rosay, W.~Maas, J.~Struppe,
  C.~Suiter, J.~Ahn, I.~L. Byeon, W.~T. Franks, M.~Orwick-Rydmark,
  A.~Bertarello, H.~Oschkinat, A.~Lesage, G.~Pintacuda, A.~M. Gronenborn,
  T.~Polenova, \textit{Dynamic Nuclear Polarization Enhanced MAS NMR
  Spectroscopy for Structural Analysis of HIV-1 Protein Assemblies}.
\newblock {\it J. Phys. Chem. B\/} {\bf 120}, 329-339 (2016).

\bibitem{Lesage2010}
A.~Lesage, M.~Lelli, D.~Gajan, M.~A. Caporini, V.~Vitzthum, P.~Mi{\'e}ville,
  J.~Alauzun, A.~Roussey, C.~Thieuleux, A.~Mehdi, G.~Bodenhausen, C.~Coperet,
  L.~Emsley, \textit{Surface Enhanced NMR Spectroscopy by Dynamic Nuclear
  Polarization}.
\newblock {\it J. Am. Chem. Soc.\/} {\bf 132}, 15459-15461 (2010).

\bibitem{Rossini2014}
A.~J. Rossini, C.~M. Widdifield, A.~Zagdoun, M.~Lelli, M.~Schwarzw{\"a}lder,
  C.~Cop{\'e}ret, A.~Lesage, L.~Emsley, \textit{Dynamic Nuclear Polarization
  Enhanced NMR Spectroscopy for Pharmaceutical Formulations}.
\newblock {\it J. Am. Chem. Soc.\/} {\bf 136}, 2324-2334 (2014).

\bibitem{Berruyer2017}
P.~Berruyer, M.~Lelli, M.~P. Conley, D.~L. Silverio, C.~M. Widdifield,
  G.~Siddiqi, D.~Gajan, A.~Lesage, C.~Cop{\'e}ret, L.~Emsley,
  \textit{Three-Dimensional Structure Determination of Surface Sites}.
\newblock {\it J. Am. Chem. Soc.\/} {\bf 139}, 849-855 (2017).

\bibitem{Golman2003}
K.~Golman, J.~H. Ardenkjær-Larsen, J.~S. Petersson, S.~M{\aa}nsson,
  I.~Leunbach, \textit{Molecular Imaging with Endogenous Substances}.
\newblock {\it Proc. Natl. Acad. Sci. U.S.A\/} {\bf 100}, 10435-10439 (2003).

\bibitem{Krishna2002}
M.~C. Krishna, S.~English, K.~Yamada, J.~Yoo, R.~Murugesan, N.~Devasahayam,
  J.~A. Cook, K.~Golman, J.~H. Ardenkjaer-Larsen, S.~Subramanian, J.~B.
  Mitchell, \textit{Overhauser Enhanced Magnetic Resonance Imaging for Tumor
  Oximetry: Coregistration of Tumor Anatomy and Tissue Oxygen Concentration}.
\newblock {\it Proc. Natl. Acad. Sci. U.S.A\/} {\bf 99}, 2216-2221 (2002).

\bibitem{Lee2013}
Y.~Lee, G.~S. Heo, H.~Zeng, K.~L. Wooley, C.~Hilty, \textit{Detection of Living
  Anionic Species in Polymerization Reactions Using Hyperpolarized NMR}.
\newblock {\it J. Am. Chem. Soc\/} {\bf 135}, 4636-4639 (2013).

\bibitem{Ji2017}
X.~Ji, A.~Bornet, B.~Vuichoud, J.~Milani, D.~Gajan, A.~J. Rossini, L.~Emsley,
  G.~Bodenhausen, S.~Jannin, \textit{Transportable Hyperpolarized Metabolites}.
\newblock {\it Nat. Commun.\/} {\bf 8}, 13975 (2017).

\bibitem{Chen2013}
H.~Chen, M.~Ragavan, C.~Hilty, \textit{Protein Folding Studied by Dissolution
  Dynamic Nuclear Polarization}.
\newblock {\it Angew. Chem. Int. Ed.\/} {\bf 52}, 9192-9195 (2013).

\bibitem{Nelson2013}
S.~J. Nelson, J.~Kurhanewicz, D.~B. Vigneron, P.~E.~Z. Larson, A.~L. Harzstark,
  M.~Ferrone, M.~van Criekinge, J.~W. Chang, R.~Bok, I.~Park, G.~Reed,
  L.~Carvajal, E.~J. Small, P.~Munster, V.~K. Weinberg, J.~H.
  Ardenkjaer-Larsen, A.~P. Chen, R.~E. Hurd, L.~Odegardstuen, F.~J. Robb,
  J.~Tropp, J.~A. Murray, \textit{Metabolic Imaging of Patients with Prostate
  Cancer Using Hyperpolarized [1-13C] Pyruvate}.
\newblock {\it Sci. Transl. Med.\/} {\bf 5}, 198ra108-198ra108 (2013).

\bibitem{Mamin2003}
H.~J. Mamin, R.~Budakian, B.~W. Chui, D.~Rugar, \textit{Detection and
  Manipulation of Statistical Polarization in Small Spin Ensembles}.
\newblock {\it Phys. Rev. Lett.\/} {\bf 91}, 207604 (2003).

\bibitem{Herzog2014}
B.~E. Herzog, D.~Cadeddu, F.~Xue, P.~Peddibhotla, M.~Poggio, \textit{Boundary
  Between the Thermal and Statistical Polarization Regimes in a Nuclear Spin
  Ensemble}.
\newblock {\it Appl. Phys. Lett.\/} {\bf 105}, 043112 (2014).

\bibitem{Staudenmaier2023}
N.~Staudenmaier, A.~Vijayakumar-Sreeja, G.~Genov, D.~Cohen, C.~Findler,
  J.~Lang, A.~Retzker, F.~Jelezko, S.~Oviedo-Casado, \textit{Optimal Sensing
  Protocol for Statistically Polarized Nano-NMR with NV Centers}.
\newblock {\it Phys. Rev. Lett.\/} {\bf 131}, 150801 (2023).

\bibitem{Lumata2013}
L.~Lumata, Z.~Kovacs, A.~D. Sherry, C.~Malloy, S.~Hill, J.~van Tol, L.~Yu,
  L.~Song, M.~E. Merritt, \textit{Electron Spin Resonance Studies of Trityl
  OX063 at a Concentration Optimal for DNP}.
\newblock {\it Phys. Chem. Chem. Phys.\/} {\bf 15}, 9800-9807 (2013).

\bibitem{Can2017}
T.~V. Can, R.~T. Weber, J.~J. Walish, T.~M. Swager, R.~G. Griffin,
  \textit{Ramped-amplitude NOVEL}.
\newblock {\it J. Chem. Phys.\/} {\bf 146}, 154204 (2017).

\bibitem{Corzilius2020}
B.~Corzilius, \textit{High-Field Dynamic Nuclear Polarization}.
\newblock {\it Annu. Rev. Phys. Chem.\/} {\bf 71}, 143-170 (2020).

\bibitem{Mathies2016}
G.~Mathies, S.~Jain, M.~Reese, R.~G. Griffin, \textit{Pulsed Dynamic Nuclear
  Polarization with Trityl Radicals}.
\newblock {\it J. Phys. Chem. Lett.\/} {\bf 7}, 111-116 (2016).

\bibitem{Issac2016}
C.~E. Issac, C.~M. Gleave, P.~T. Nasr, H.~L. Nguyen, E.~A. Curley, J.~L. Yoder,
  E.~W. Moore, L.~Chen, J.~A. Marohn, \textit{Dynamic Nuclear Polarization in a
  Magnetic Resonance Force Microscope Experiment}.
\newblock {\it Phys. Chem. Chem. Phys.\/} {\bf 18}, 8806-8819 (2016).

\bibitem{Chen2016}
H.~Chen, A.~G. Maryasov, O.~Y. Rogozhnikova, D.~V. Trukhin, V.~M. Tormyshev,
  M.~K. Bowman, \textit{Electron Spin Dynamics and Spin–Lattice Relaxation of
  Trityl Radicals in Frozen Solutions}.
\newblock {\it Phys. Chem. Chem. Phys.\/} {\bf 18}, 24954-24965 (2016).

\bibitem{Kolkowitz2015}
S.~Kolkowitz, A.~Safira, A.~A. High, R.~C. Devlin, S.~Choi, Q.~P.
  Unterreithmeier, D.~Patterson, A.~S. Zibrov, V.~E. Manucharyan, H.~Park,
  M.~D. Lukin, \textit{Probing Johnson Noise and Ballistic Transport in Normal
  Metals with a Single-Spin Qubit}.
\newblock {\it Science\/} {\bf 347}, 1129-1132 (2015).

\bibitem{Wang2006}
R.~P. Wang, \textit{Defects in Silicon Nanowires}.
\newblock {\it Appl. Phys. Lett.\/} {\bf 88}, 142104 (2006).

\bibitem{Baumer2004}
A.~Baumer, M.~Stutzmann, M.~S. Brandt, F.~C. Au, S.~T. Lee,
  \textit{Paramagnetic Defects of Silicon Nanowires}.
\newblock {\it Appl. Phys. Lett.\/} {\bf 85}, 943-945 (2004).

\bibitem{Brower1986}
K.~L. Brower, T.~J. Headley, \textit{Dipolar Interactions Between Dangling
  Bonds at the (111) Si-SiO$_{2}$ Interface}.
\newblock {\it Phys. Rev. B\/} {\bf 34}, 3610--3619 (1986).

\bibitem{Nouwen2000}
B.~Nouwen, A.~Stesmans, \textit{Dependence of Strain at the (111) Si/SiO$_{2}$
  Interface on Interfacial Si Dangling-Bond Concentration}.
\newblock {\it Mater. Sci. Eng., A\/} {\bf 288}, 239-243 (2000).

\bibitem{Rurali2010}
R.~Rurali, \textit{Colloquium: Structural, Electronic, and Transport Properties
  of Silicon Nanowires}.
\newblock {\it Rev. Mod. Phys.\/} {\bf 82}, 427--449 (2010).

\bibitem{Stesmans1989}
A.~Stesmans, \textit{The .Si Identical to Si$_3$ Defect at Various (111)Si/SiO2
  and (111) Si/Si$_3$N$_4$ Interfaces}.
\newblock {\it Semicond. Sci. Technol.\/} {\bf 4}, 1000 (1989).

\bibitem{Bloembergen1949}
N.~Bloembergen, \textit{On the Interaction of Nuclear Spins in a Crystalline
  Lattice}.
\newblock {\it Physica\/} {\bf 15}, 386-426 (1949).

\bibitem{Dementyev2008}
A.~E. Dementyev, D.~G. Cory, C.~Ramanathan, \textit{Dynamic Nuclear
  Polarization in Silicon Microparticles}.
\newblock {\it Phys. Rev. Lett.\/} {\bf 100}, 127601 (2008).

\bibitem{Poggio2007}
M.~Poggio, C.~L. Degen, H.~J. Mamin, D.~Rugar, \textit{Feedback Cooling of a
  Cantilever's Fundamental Mode below 5~mK}.
\newblock {\it Phys. Rev. Lett\/} {\bf 99}, 017201 (2007).

\end{thebibliography}


\noindent \textbf{Acknowledgements:} 
The University of Waterloo's QNFCF facility was used for this work. This infrastructure would not be possible without the significant contributions of CFREF-TQT, CFI, Industry Canada, the Ontario Ministry of Research and Innovation and Mike and Ophelia Lazaridis. Their support is gratefully acknowledged.
We would like to thank D. Akhmetzyanov and T. W. Borneman for helpful discussions and D. G. Cory for providing the trityl-OX063 samples, and useful comments.
\\
\noindent \textbf{Funding:} This work was undertaken thanks in part to funding from the Canada First Research Excellence
Fund (CFREF), and the Natural Sciences and Engineering Research Council of Canada (NSERC). \\
\noindent \textbf{Author Contributions:} Conceptualization: R.B. Measurements and data analysis: S.T., P.P. and N.S. Simulations: N.S., D.T., P.P. and S.T.  SiNW growth: P.S. and A.J. Sample preparation: P.P. Writing (original draft): S.T. and R.B. Writing (review and edits): S.T., P.P., N.S, D.T. and R.B. Supervision: R.B. Resources: R.B.\\
\noindent \textbf{Competing Interests:} The authors declare that they have no competing financial interests.\\
\noindent \textbf{Data and materials availability:} All data needed to evaluate the conclusions in the paper are present in the paper and/or the Supplementary Materials.

\clearpage
\clearpage
\includepdf[pages={1-}]{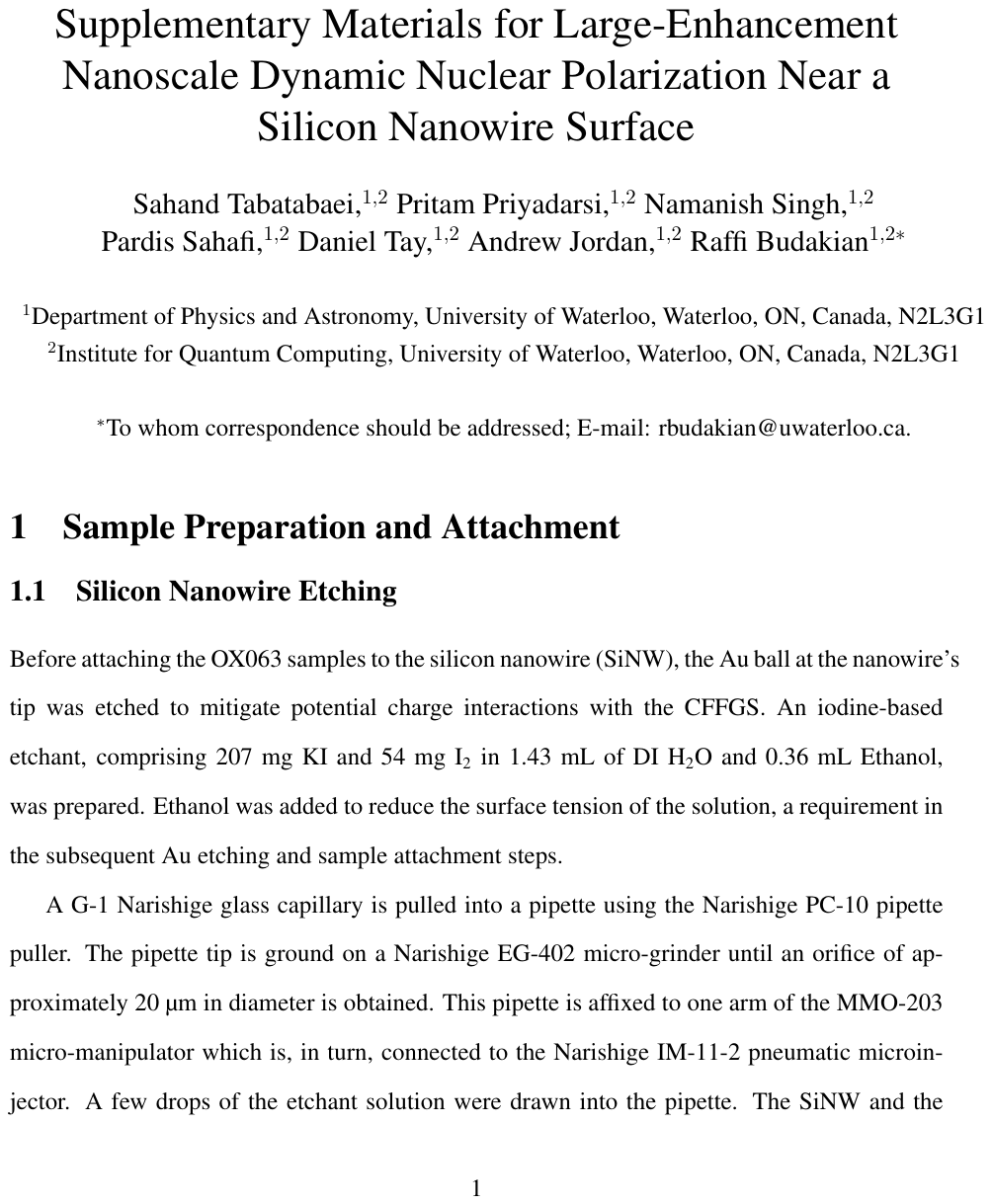}
\end{document}